\documentclass[preprintnumbers,amsmath,amssymbm,preprint]{revtex4}
\usepackage{epsfig}
\usepackage{graphicx}

\begin{document}
\title{Resonance spectrum of near-extremal
Kerr black holes in the eikonal limit}
\author{Shahar Hod}
\address{The Ruppin Academic Center, Emeq Hefer 40250, Israel}
\address{ }
\address{The Hadassah Institute, Jerusalem 91010, Israel}
\date{\today}

\begin{abstract}
\ \ \ The fundamental resonances of rapidly rotating Kerr black
holes in the eikonal limit are derived analytically. We show that
there exists a critical value,
$\mu_c=\sqrt{{{15-\sqrt{193}}\over{2}}}$, for the dimensionless
ratio $\mu\equiv m/l$ between the azimuthal harmonic index $m$ and
the spheroidal harmonic index $l$ of the perturbation mode, above
which the perturbations become long lived. In particular, it is
proved that above $\mu_c$ the imaginary parts of the quasinormal
frequencies scale like the black-hole temperature:
$\omega_I(n;\mu>\mu_c)=2\pi T_{BH}(n+{1\over 2})$. This implies that
for perturbations modes in the interval $\mu_c<\mu\leq 1$, the
relaxation period $\tau\sim 1/\omega_I$ of the black hole becomes
extremely long as the extremal limit $T_{BH}\to 0$ is approached. A
generalization of the results to the case of scalar quasinormal
resonances of near-extremal Kerr-Newman black holes is also
provided. In particular, we prove that only black holes that rotate
fast enough (with $M\Omega\geq {2\over 5}$, where $M$ and $\Omega$
are the black-hole mass and angular velocity, respectively) possess
this family of remarkably long-lived perturbation modes.
\end{abstract}
\bigskip
\maketitle


\section{Introduction}

The response of a black hole to external perturbations is
characterized by `quasinormal ringing', damped (complex)
oscillations with a discrete frequency spectrum (see
\cite{Nollert1,Ber1} for excellent reviews and detailed lists of
references). This implies that radiative perturbations of the
black-hole spacetime fade away over time in a manner reminiscent of
the last pure dying tones of a ringing bell \cite{Press}. This
characteristic decay of black-hole perturbations is in accord with
the no-hair conjecture \cite{Whee} which asserts that the external
field of a perturbed black hole should relax into a Kerr-Newman
spacetime, characterized solely by three observable (conserved)
parameters: the black-hole mass, charge, and angular momentum.

This relaxation phase in the dynamics of perturbed black holes is
characterized by a temporal decay of the perturbation fields of the
form $e^{-i\omega t}$, where the characteristic black-hole resonance
frequencies are complex numbers ($\omega=\omega_R-i\omega_I$ with
$\omega_I\geq 0$) that depend on the black-hole physical parameters
. These damped oscillations are then followed by late-time decaying
tails that depend on the asymptotic properties of the spacetime
\cite{Tails1,Tails2}.

The black-hole quasinormal modes (QNMs) correspond to solutions of
the perturbations equations with the physical boundary conditions of
purely outgoing waves at spatial infinity and purely ingoing waves
crossing the black-hole horizon \cite{Detw}. These boundary
conditions single out a discrete and infinite family of black-hole
resonances \cite{Noteres} $\{\omega(n;m,l)\}$, where $l$ and $m$ are
the multipolar indexes of the angular eigenfunctions which
correspond to the QNMs [see Eqs. (\ref{Eq2})-(\ref{Eq3}) below]. The
resonance parameter $n$ is a non-negative integer which
characterizes the overtone number.

Quasinormal resonances are expected to play a prominent role in
gravitational radiation emitted by a variety of astrophysical
scenarios involving black holes. Given the fact that these damped
oscillations are the characteristic `sound' of the black hole
itself, they have attracted much attention from both physicists and
mathematicians. In particular, the spectrum of black-hole QNMs is of
great importance from both the theoretical \cite{HodPRL,Gary} and
astrophysical points of view \cite{Nollert1,Ber1}. These black-hole
characteristic oscillations provide a direct way of identifying the
black-hole parameters. This fact has motivated a flurry of research
during the last four decades aiming to compute the resonance
spectrum of various types of black holes \cite{Nollert1,Ber1}.

It turns out that for fixed values of the multipolar indexes $m$ and
$l$ there exist an infinite number of QNMs, characterizing
oscillations with decreasing relaxation times (increasing imaginary
part), see \cite{Leaver,Gary,Noll2,HodPRL,KeshHod} and references
therein. The mode with the smallest imaginary part $ -$ known as the
fundamental mode $ -$ determines the characteristic dynamical
timescale for generic perturbations to decay
\cite{GlaAnd2,Hod1,Hod2,Hod3}.

It is worth emphasizing that in most situations of physical interest
the spectrum of QNMs must be computed {\it numerically} by solving
the black-hole perturbations equations supplemented by the
appropriate physical boundary conditions \cite{Notephys}. However,
Mashhoon \cite{Mash} has developed an {\it analytical} technique for
calculating the equatorial QNMs of rotating Kerr black holes in the
eikonal (geometric-optics) limit $l=m\gg 1$, see also
\cite{Goeb,CarMir,Dolan,Massfr}.

Recently, Yang et. al. \cite{Yang} have generalized the large-$l$
analysis to include non-equatorial modes with $l\neq m$. The
analysis of Yang et. al. \cite{Yang} is remarkably elegant and
intuitive. Yet, their final expressions for the quasinormal
frequencies are rather complicated: see Eqs. (2.35)-(2.36) of
\cite{Yang} for the real parts of the frequencies and Eqs. (2.36)
and (2.40) of \cite{Yang} for the imaginary parts of the
frequencies. These equations must then be solved {\it numerically}
in order to obtain the corresponding quasinormal frequencies; the
numerical values of these quasinormal frequencies are presented in
Fig. 3 (real parts) and Fig. 5 (imaginary parts) of \cite{Yang}.

One of the most remarkable conclusions of \cite{Yang} is that
near-extremal Kerr black holes are characterized by a significant
fraction of QNMs that have nearly {\it zero} imaginary part (see
also \cite{Hod1,Hod2,Hod3}). These resonances thus correspond to
black-hole perturbations which may survive for relatively long times
as compared to the dynamical timescale set by the mass of the black
hole. In particular, it was observed numerically in \cite{Yang} that
such long-lived modes exist for near-extremal black holes in the
finite interval
\begin{equation}\label{Eq1}
\mu_c\equiv 0.74\lesssim {m\over l}\leq 1\ \ \ \text{for}\ \ \ l\gg
1\ .
\end{equation}
Thus, not only for equatorial modes with $l=m$ \cite{Mash} does
$\omega_I$ vanish in the extremal limit! Below we shall provide a
fully {\it analytical} explanation for this phenomena. Furthermore,
we shall obtain an analytical expression for the exact value of the
critical ratio $\mu_c$ above which the long-lived modes appear.

\section{Description of the system}

In order to determine the black-hole quasinormal resonances we shall
study the scattering of massless fields in the Kerr black-hole
spacetime. The dynamics of a perturbation field $\Psi$ in the
rotating Kerr spacetime is governed by the Teukolsky master equation
\cite{Teu}. As we shall show below, the Teukolsky equation is
amenable to an {\it analytical} treatment in the near-extremal limit
$(M^2-a^2)^{1/2}\ll a\lesssim M$ (we use units in which
$G=c=\hbar=1$), where $M$ and $a$ are the black-hole mass and
angular momentum per unit mass, respectively.

One may decompose the field as
\begin{equation}\label{Eq2}
\Psi_{slm}(t,r,\theta,\phi)=e^{im\phi}S_{slm}(\theta;a\omega)\psi_{slm}(r)e^{-i\omega
t}\ ,
\end{equation}
where $(t,r,\theta,\phi)$ are the Boyer-Lindquist coordinates,
$\omega$ is the (conserved) frequency of the mode, $l$ is the
spheroidal harmonic index, and $m$ is the azimuthal harmonic index
with $-l\leq m\leq l$. The parameter $s$ is called the spin weight
of the field, and is given by $s=\pm 2$ for gravitational
perturbations, $s=\pm 1$ for electromagnetic perturbations, $s=\pm
{1\over 2}$ for massless neutrino perturbations, and $s=0$ for
scalar perturbations \cite{Teu}. (We shall henceforth omit the
indices $s,l,m$ for brevity.) With the decomposition (\ref{Eq2}),
$\psi$ and $S$ obey radial and angular equations, both of confluent
Heun type \cite{Teu,Heun,Flam,Fiz1}, coupled by a separation
constant $A(a\omega)$.

The angular functions $S(\theta;a\omega)$ are the spin-weighted
spheroidal harmonics which are solutions of the angular equation
\cite{Teu,Heun,Flam}
\begin{equation}\label{Eq3}
{1\over {\sin\theta}}{\partial \over
{\partial\theta}}\Big(\sin\theta {{\partial
S}\over{\partial\theta}}\Big)+\Big[a^2\omega^2\cos^2\theta-2a\omega
s\cos\theta-{{(m+s\cos\theta)^2}\over{\sin^2\theta}}+s+A\Big]S=0\ .
\end{equation}
The angular functions are required to be regular at the poles
$\theta=0$ and $\theta=\pi$. These boundary conditions pick out a
discrete set of eigenvalues $_sA_{lm}$ labeled by the integers $m$
and $l$. [In the $a\omega\ll 1$ limit these angular functions become
the familiar spin-weighted spherical harmonics with the
corresponding angular eigenvalues $A=l(l+1)-s(s+1)+O(a\omega)$.] The
angular equation (\ref{Eq3}) can be solved analytically in the $l\gg
1$ limit (with $\omega_R\gg\omega_I$) to yield \cite{Yang}
\begin{equation}\label{Eq4}
A=l(l+1)-{1\over 2}a^2\omega^2_R(1-\mu^2)+O(1)\ ,
\end{equation}
where here
\begin{equation}\label{Eq5}
\mu\equiv {m\over l}\
\end{equation}
is the dimensionless ratio between the azimuthal harmonic index $m$
and the spheroidal harmonic index $l$.

The radial Teukolsky equation is given by \cite{Teu}
\begin{equation}\label{Eq6}
\Delta^{-s}{{d}
\over{dr}}\Big(\Delta^{s+1}{{d\psi}\over{dr}}\Big)+\Big[{{K^2-2is(r-M)K}\over{\Delta}}
-a^2\omega^2+2ma\omega-A+4is\omega
r\Big]\psi=0\ ,
\end{equation}
where $\Delta\equiv r^2-2Mr+a^2$ and $K\equiv (r^2+a^2)\omega-am$.
The zeroes of $\Delta$, $r_{\pm}=M\pm (M^2-a^2)^{1/2}$, are the
black hole (event and inner) horizons.

For the problem of wave-scattering in a black-hole spacetime one
should impose physical boundary conditions of purely ingoing waves
at the black-hole horizon and a mixture of both ingoing and outgoing
waves at spatial infinity (these correspond to incident and
scattered waves, respectively). That is,
\begin{equation}\label{Eq7}
\psi \sim
\begin{cases}
e^{-i\omega y}+{\mathcal{R}}(\omega)e^{i \omega y} & \text{ as }
r\rightarrow\infty\ \ (y\rightarrow \infty)\ ; \\
{\mathcal{T}}(\omega)e^{-i (\omega-m\Omega)y} & \text{ as }
r\rightarrow r_+\ \ (y\rightarrow -\infty)\ ,
\end{cases}
\end{equation}
where the ``tortoise" radial coordinate $y$ is defined by
$dy=[(r^2+a^2)/\Delta]dr$. Here $\Omega$ is the angular velocity of
the black hole [see Eq. (\ref{Eq8}) below]. The coefficients ${\cal
T}(\omega)$ and ${\cal R}(\omega)$ are the transmission and
reflection amplitudes for a wave incident from infinity. They
satisfy the usual probability conservation equation $|{\cal
T}(\omega)|^2+|{\cal R}(\omega)|^2=1$.

\section{The quasinormal resonances}

The discrete family of quasinormal frequencies describes the
scattering resonances of the fields in the black-hole spacetime.
These resonances correspond to poles of the transmission and
reflection amplitudes. (The pole structure reflects the fact that
the QNMs correspond to purely outgoing waves at spatial infinity.)
These resonances determine the ringdown response of a black hole to
external perturbations. As we shall now show, the spectrum of
quasinormal frequencies can be studied {\it analytically} in the
near-extremal limit $a\to M$, see also \cite{Hod1,Hod2,Hod3}.

Teukolsky and Press \cite{TeuPre} and also Starobinsky and Churilov
\cite{StaChu} have studied the black-hole scattering problem in the
double limit $a\to M$ ($T_{BH}\to 0$) and $\omega\to m\Omega$, where
\begin{equation}\label{Eq8}
T_{BH}\equiv {{r_+-r_-}\over{4\pi(r^2_++a^2)}}\ \ \ ; \ \ \
\Omega\equiv {{a}\over{r^2_++a^2}}
\end{equation}
are the black-hole temperature and angular momentum, respectively.
Detweiler \cite{Det} then used the analysis of \cite{TeuPre,StaChu}
to obtain a resonance condition for near-extremal Kerr black holes.
It is convenient to define a set of dimensionless variables:
\begin{equation}\label{Eq9}
\sigma\equiv {{r_+-r_-}\over{r_+}}\ \ ;\ \ \tau\equiv
M(\omega-m\Omega)\ \ ;\ \ \hat\omega\equiv \omega r_+\  ,
\end{equation}
in terms of which the resonance condition obtained in \cite{Det} for
$\sigma<<1$ and $\tau<<1$ is:
\begin{eqnarray}\label{Eq10}
-{{\Gamma(2i\delta)\Gamma(1+2i\delta)\Gamma(1/2+s-2i\hat\omega-i\delta)\Gamma(1/2-s-2i\hat\omega-i\delta)}\over
{\Gamma(-2i\delta)\Gamma(1-2i\delta)\Gamma(1/2+s-2i\hat\omega+i\delta)\Gamma(1/2-s-2i\hat\omega+i\delta)}}
\nonumber
\\
= (-2i\hat\omega\sigma)^{2i\delta}
{{\Gamma(1/2+2i\hat\omega+i\delta-4i\tau/\sigma)}\over{\Gamma(1/2+2i\hat\omega-i\delta-4i\tau/\sigma)}}\
,
\end{eqnarray}
where
\begin{equation}\label{Eq11}
\delta^2\equiv 4\hat\omega^2-1/4-A-a^2\omega^2+2ma\omega\  .
\end{equation}
We shall assume without loss of generality that $\Re\delta\geq 0$.
Taking cognizance of Eq. (\ref{Eq4}), one finds
\begin{equation}\label{Eq12}
\delta=l\times {\cal F}(\mu)+O(1)
\end{equation}
for near-extremal Kerr black holes in the eikonal limit $l\gg 1$,
where
\begin{equation}\label{Eq13}
{\cal F}(\mu)\equiv\sqrt{-1+{15\over 8}\mu^2-{1\over8}\mu^4}\ .
\end{equation}
Here we have used the relations $a\simeq r_+\simeq M$ and
$\omega\simeq m\Omega\simeq m/2M$ for near-extremal Kerr black
holes. Note that the function ${\cal F}(\mu)$ is real and positive
in the interval
\begin{equation}\label{Eq14}
\mu_c\equiv\sqrt{{{15-\sqrt{193}}\over{2}}}<\mu\leq 1\  .
\end{equation}

The resonance condition (\ref{Eq10}) can be solved analytically in
the regime $\sigma\ll 1$ with $\omega\simeq m\Omega$
\cite{Hod1,Hod2,Hod3}. The l.h.s of it has a well defined limit as
$a\to M$ and $\omega\to m\Omega$. We denote that limit by ${\cal
L}$. Given the fact that $\delta\simeq l{\cal F}\gg 1$ is purely
real and large for $l\gg 1$ in the interval $\mu_c<\mu\leq 1$ [see
Eq. (\ref{Eq14})], one finds
$(-i)^{-2i\delta}=e^{(-i{{\pi}\over{2}})(-2i\delta)}=e^{-\pi l{\cal
F}}\ll 1$, which implies
$\epsilon\equiv(-2i\hat\omega\sigma)^{-2i\delta}\ll 1$. Thus, a
consistent solution of the resonance condition, Eq. (\ref{Eq10}),
may be obtained if
$1/\Gamma(1/2+2i\hat\omega-i\delta-4i\tau/\sigma)=O(\epsilon)$.
Suppose
\begin{equation}\label{Eq15}
1/2+2i\hat\omega-i\delta-4i\tau/\sigma=-n+\eta\epsilon+O(\epsilon^2)\
,
\end{equation}
where $n\geq 0$ is a non-negative integer and $\eta$ is an unknown
constant to be determined below. Then one has
\begin{equation}\label{Eq16}
\Gamma(1/2+2i\hat\omega-i\delta-4i\tau/\sigma)\simeq\Gamma(-n+\eta\epsilon)\simeq
(-n)^{-1}\Gamma(-n+1+\eta\epsilon)\simeq\cdots\simeq [(-1)^n
n!]^{-1}\Gamma(\eta\epsilon)\  ,
\end{equation}
where we have used the relation $\Gamma(z+1)=z\Gamma(z)$
\cite{Abram}. Next, using the series expansion
$1/\Gamma(z)=\sum_{k=1}^{\infty} c_k z^k$ with $c_1=1$ \cite{Abram},
one obtains
\begin{equation}\label{Eq17}
1/\Gamma(1/2+2i\hat\omega-i\delta-4i\tau/\sigma)=(-1)^n
n!\eta\epsilon+O(\epsilon^2)\  .
\end{equation}
Substituting (\ref{Eq17}) into the resonance condition (\ref{Eq10})
one finds $\eta={\cal L}/[(-1)^n n!\Gamma(-n+2i\delta)]$.

Finally, substituting $4\tau/\sigma=(\omega-m\Omega)/2\pi T_{BH}$,
$2i\hat\omega=im+O(mMT_{BH})$ and $\delta=l{\cal F}+O(1)$ for
$\omega=m\Omega+O(mT_{BH})$ into Eq. (\ref{Eq15}), one obtains the
resonance condition
\begin{equation}\label{Eq18}
(\omega-m\Omega)/2\pi T_{BH}=i[-n+\eta\epsilon-1/2]+m-l{\cal F}\ .
\end{equation}
Thus, the spectrum of black-hole quasinormal resonances in the
eikonal limit $l\gg 1$ within the interval $\mu_c<\mu\leq 1$ is
described by the compact analytical formula \cite{Hodol}
\begin{equation}\label{Eq19}
\omega(n;\mu>\mu_c)=m\Omega+2\pi T_{BH}\big[(1-{\cal
F}/\mu)m-i(n+{1\over 2})\big]+O(MT^2_{BH})\ \ \ ; \ \ \ n=0,1,2,...\
,
\end{equation}
where the dimensionless function ${\cal F}(\mu)$ is defined in
(\ref{Eq13}). Note that the spectrum (\ref{Eq19}) corresponds to
black-hole perturbation modes with relaxation times $\tau\sim
1/\omega_I$ that become extremely long as the extremal limit
$T_{BH}\to 0$ is approached. We also note that the numerically
computed value $\mu_c\simeq 0.74$ \cite{Yang} is astonishingly close
to the analytical expression (\ref{Eq14}) for the critical ratio
$\mu_c$.

\section{Scalar QNMs of Kerr-Newman black holes}

Our analysis can readily be generalized to the case of scalar
quasinormal resonances of charged and rotating Kerr-Newman (KN)
black holes \cite{Ber1,Hart} of mass $M$, angular momentum per unit
mass $a$, and charge $Q$. Substituting the relations $r_+\simeq M$
and $\omega\simeq m\Omega\simeq ma/(M^2+a^2)$ \cite{Noterp} into
Eqs. (\ref{Eq4}), (\ref{Eq11}) and (\ref{Eq12}), one finds
\begin{equation}\label{Eq20}
{\cal F}(\mu;\alpha)\equiv \sqrt{-1+{{6\alpha^2(1+{1\over
4}\alpha^2)}\over{(1+\alpha^2)^2}}\mu^2-{{\alpha^4}\over{2(1+\alpha^2)^2}}\mu^4}\
\end{equation}
for near-extremal KN black holes, where
\begin{equation}\label{Eq21}
\alpha\equiv a/M
\end{equation}
is the rescaled (dimensionless) angular momentum of the black hole.
Note that Eq. (\ref{Eq20}) is merely a generalization of
(\ref{Eq13}) and reduces to it in the limit $a\to M$ ($\alpha\to
1$). The critical ratio $\mu_c(\alpha)$ is obtained from the
limiting case ${\cal F}=0$. One obtains
\begin{equation}\label{Eq22}
\mu_c(\alpha)={1\over{2\alpha}}\sqrt{6\alpha^2+24-\sqrt{4\alpha^4+224\alpha^2+544}}\
\end{equation}
for near extremal KN black holes.

Using an analysis along the same lines as before, one finds that the
scalar quasinormal mode spectrum of near-extremal KN black holes in
the eikonal limit $l\gg 1$
is described by the formula
\begin{equation}\label{Eq23}
\omega(n;\mu>\mu_c(\alpha))=m\Omega+2\pi T_{BH}\big[(1-{\cal
F(\mu;\alpha)}/\mu)m-i(n+{1\over 2})\big]+O(MT^2_{BH})\ \ \ ; \ \ \
n=0,1,2,...\ ,
\end{equation}
where the functions ${\cal F}(\mu;\alpha)$ and $\mu_c(\alpha)$ are
defined in Eqs. (\ref{Eq20}) and (\ref{Eq22}), respectively.

Inspection of Eq. (\ref{Eq22}) reveals that the interval
$\mu_c(\alpha)<\mu\leq 1$ shrinks as the ratio $\alpha$ decreases
[For example, for near-extremal black holes with $\alpha=0.75$ one
finds $\mu_c(\alpha=0.75)\simeq 0.807$ instead of the value
$\mu_c(\alpha=1)=\sqrt{{{15-\sqrt{193}}\over{2}}}\simeq 0.744$.]
Remarkably, we find that the condition $\mu_c(\alpha)<\mu\leq 1$ can
only be satisfied by rotating black holes in the interval
\begin{equation}\label{Eq24}
{1\over 2}\leq\alpha\leq 1\  .
\end{equation}
One therefore concludes that, within the eikonal approximation, only
KN black holes that rotate fast enough ($a\geq {1\over 2}M$, or
equivalently $M\Omega\geq {2\over 5}$) can be characterized by
relaxation periods $\tau\sim 1/\omega_I$ which become infinitely
long as the extremal limit $T_{BH}\to 0$ is approached.

\section{Summary}

In summary, the quasinormal mode spectrum of near-extremal rotating
Kerr black holes was studied {\it analytically} within the eikonal
approximation $l\gg 1$. It was shown that the fundamental resonances
can be expressed in terms of the black-hole physical parameters: the
temperature $T_{BH}$ and the angular velocity $\Omega$. In
particular, we have proved the existence of a critical value,
$\mu_c$, for the dimensionless ratio $\mu\equiv m/l$ between the
azimuthal harmonic index $m$ and the spheroidal harmonic index $l$
of the perturbation mode, above which the perturbations become {\it
long lived} -- for modes in the interval $\mu_c<\mu\leq 1$ the
imaginary parts of the quasinormal frequencies scale like the
black-hole temperature: $\omega_I(n;\mu>\mu_c)=2\pi T_{BH}(n+{1\over
2})$. For these perturbation modes the relaxation period $\tau\sim
1/\omega_I$ of the black hole becomes extremely long as the extremal
limit $T_{BH}\to 0$ is approached. Our analytical expression
(\ref{Eq14}) for the critical ratio,
$\mu_c=\sqrt{{{15-\sqrt{193}}\over{2}}}$, is remarkably close to the
numerically computed \cite{Yang} value $\mu_c\simeq 0.74$.

We have also generalized the results to the case of charged and
rotating Kerr-Newman black holes, proving that the interval
$\mu_c(\alpha)<\mu\leq 1$ in which the long-lived modes exist
shrinks as the dimensionless ratio $\alpha\equiv a/M$ decreases. In
particular, we have shown that only near-extremal black holes that
rotate fast enough (with $M\Omega\geq {2\over 5})$ possess this
family of extremely long-lived perturbation modes.

\bigskip
\noindent
{\bf ACKNOWLEDGMENTS}
\bigskip

This research is supported by the Carmel Science Foundation. I thank
O. Hod, Yael Oren, Arbel M. Ongo and Ayelet B. Lata for helpful
discussions.



\begin{thebibliography}{99}

\bibitem{Nollert1} H. P. Nollert, Class. Quantum Grav. {\bf 16}, R159 (1999).

\bibitem{Ber1} E. Berti, V. Cardoso and A. O. Starinets, Class. Quant. Grav. {\bf 26}, 163001
(2009).

\bibitem{Press} W. H. Press, Astrophys. J. {\bf 170}, L105 (1971);
V. de la Cruz, J. E. Chase and W. Israel, Phys. Rev. Lett. {\bf 24},
423 (1970); C.V. Vishveshwara, Nature {\bf 227}, 936 (1970); M.
Davis, R. Ruffini, W. H. Press and R. H. Price, Phys. Rev. Lett.
{\bf 27}, 1466 (1971).

\bibitem{Whee} R. Ruffini and J. A. Wheeler, Phys. Today {\bf 24}, 30
(1971).

\bibitem{Tails1} R.H. Price, Phys. Rev. D {\bf 5}, 2419 (1972);
C. Gundlach, R.H. Price, and J. Pullin, Phys. Rev. D {\bf 49}, 883
(1994); J. Bic\'ak, Gen. Relativ. Gravitation {\bf 3}, 331 (1972).

\bibitem{Tails2} E. S. C. Ching, P. T. Leung, W. M. Suen, and K.
Young, Phys. Rev. Lett. {\bf 74}, 2414 (1995); E. S. C. Ching, P. T.
Leung, W. M. Suen, and K. Young, Phys. Rev. D {\bf 52}, 2118 (1995);
S. Hod and T. Piran, Phys. Rev. D {\bf 58}, 024017 (1998)
[arXiv:gr-qc/9712041]; S. Hod and T. Piran, Phys. Rev. D {\bf 58},
024018 (1998) [arXiv:gr-qc/9801001]; S. Hod and T. Piran, Phys. Rev.
D {\bf 58}, 044018 (1998) [arXiv:gr-qc/9801059]; S. Hod and T.
Piran, Phys. Rev. D {\bf 58}, 024019 (1998) [arXiv:gr-qc/9801060];
S. Hod, Phys. Rev. D {\bf 58}, 104022 (1998) [arXiv:gr-qc/9811032];
S. Hod, Phys. Rev. D {\bf 61}, 024033 (2000) [arXiv:gr-qc/9902072];
S. Hod, Phys. Rev. D {\bf 61}, 064018 (2000) [arXiv:gr-qc/9902073];
L. Barack, Phys. Rev. D {\bf 61}, 024026 (2000); S. Hod, Phys. Rev.
Lett. {\bf 84}, 10 (2000) [arXiv:gr-qc/9907096]; S. Hod, Phys. Rev.
D {\bf 60}, 104053 (1999) [arXiv:gr-qc/9907044]; S. Hod, Class.
Quant. Grav. {\bf 26}, 028001 (2009) [arXiv:0902.0237]; S. Hod,
Class. Quant. Grav. {\bf 18}, 1311 (2001) [arXiv:gr-qc/0008001]; S.
Hod, Phys. Rev. D {\bf 66}, 024001 (2002) [arXiv:gr-qc/0201017]; R.
J. Gleiser, R. H. Price, and J. Pullin, Class. Quant. Grav. {\bf
25}, 072001 (2008); M. Tiglio, L. E. Kidder, and S. A. Teukolsky,
Class. Quant. Grav. {\bf 25}, 105022 (2008); R. Moderski and M.
Rogatko, Phys. Rev. D {\bf 77}, 124007 (2008); X. He and J. Jing,
Nucl. Phys.B {\bf 755}, 313 (2006); H. Koyama and A. Tomimatsu,
Phys. Rev. D {\bf 65}, 084031 (2002); B. Wang, C. Molina, and E.
Abdalla, Phys. Rev. D {\bf 63}, 084001 (2001).

\bibitem{Detw} S. L. Detweiler, in Sources of Gravitational Radiation,
edited by L. Smarr (Cambridge University Press, Cambridge, England,
1979).

\bibitem{Noteres} In analogy with standard scattering theory, the QNMs can be
regarded as the scattering resonances of the black-hole spacetime.
They thus correspond to poles of the transmission and reflection
amplitudes of a standard scattering problem in a black-hole
spacetime.

\bibitem{HodPRL} S. Hod, Phys. Rev. Lett. {\bf 81}, 4293 (1998)
[arXiv:gr-qc/9812002].

\bibitem{Gary} G. T. Horowitz and V. E. Hubeny, Phys. Rev. D {\bf 62}, 024027
(2000).

\bibitem{Leaver} E. W. Leaver, Proc. R. Soc. A {\bf 402}, 285 (1985).

\bibitem{Noll2} H. P. Nollert, Phys. Rev. D {\bf 47}, 5253 (1993).

\bibitem{KeshHod} U. Keshet and S. Hod, Phys. Rev. D {\bf 76}, R061501
(2007) [arXiv:0705.1179].

\bibitem{GlaAnd2} K. Glampedakis and N. Andersson, Phys. Rev. D {\bf
64}, 104021 (2001).

\bibitem{Hod1} S. Hod, Phys. Rev. D {\bf 75}, 064013 (2007) [arXiv:gr-qc/0611004];
S. Hod, Class. and Quant. Grav. {\bf 24}, 4235 (2007)
[arXiv:0705.2306]; A. Gruzinov, arXiv:gr-qc/0705.1725; A. Pesci,
Class. Quantum Grav. {\bf 24}, 6219 (2007); S. Hod, Phys. Rev. D
{\bf 78}, 084035 (2008).

\bibitem{Hod2} S. Hod, Phys. Lett. B {\bf 666} 483 (2008) [arXiv:0810.5419];
S. Hod, Phys. Rev. D {\bf 80}, 064004 (2009) [arXiv:0909.0314]; S.
Hod, Phys. Lett. A {\bf 374}, 2901 (2010) [arXiv:1006.4439].

\bibitem{Hod3} S. Hod, Phys. Rev. D. {\bf 84}, 044046 (2011)
[arXiv:1109.4080]; S. Hod, Phys. Lett. B {\bf 710}, 349 (2012)
[arXiv:1205.5087].

\bibitem{Notephys} It is well known that astrophysically realistic black holes generally rotate
about their axis and are therefore not spherical. Thus, a realistic
model of wave dynamics in black-hole spacetimes must involve a
non-spherical background geometry with angular momentum. For a
rotating Kerr spacetime the perturbation equations are described by
the Teukolsky master equation, see Eqs. (\ref{Eq3}) and (\ref{Eq6})
below.

\bibitem{Mash} B. Mashhoon, Phys. Rev. D {\bf 31}, 290 (1985).

\bibitem{Goeb} C. J. Goebel, Astrophys. J. {\bf 172}, L95 (1972).

\bibitem{CarMir} V. Cardoso, A. S. Miranda, E. Berti, H. Witek, and V. T. Zanchin, Phys. Rev. D {\bf 79}, 064016 (2009).

\bibitem{Dolan} S. R. Dolan and A. C. Ottewill, Classical Quantum Gravity {\bf 26}, 225003 (2009);
S. R. Dolan, Phys. Rev. D {\bf 82}, 104003 (2010).

\bibitem{Massfr}  Y. D\'ecanini, A. Folacci, and B. Raffaelli, Phys. Rev. D {\bf 84}, 084035
(2011).

\bibitem{Yang} H. Yang, D. A. Nichols, F. Zhang, A. Zimmerman, Z. Zhang and Y.
Chen, arXiv:1207.4253.

\bibitem{Teu} S. A. Teukolsky, Phys. Rev. Lett. {\bf 29}, 1114 (1972);
Astrophys. J. {\bf 185}, 635 (1973).

\bibitem{Heun} A. Ronveaux, {\it Heun's differential equations}.
(Oxford University Press, Oxford, UK, 1995).

\bibitem{Flam} C. Flammer, {\it Spheroidal Wave Functions} (Stanford
University Press, Stanford, 1957).

\bibitem{Fiz1} P. P. Fiziev, e-print arXiv:0902.1277; R. S. Borissov and P. P. Fiziev, e-print arXiv:0903.3617;
P. P. Fiziev, Phys. Rev. D {\bf 80}, 124001 (2009); P. P. Fiziev,
Class. Quant. Grav. {\bf 27}, 135001 (2010).

\bibitem{TeuPre} S. A. Teukolsky and W. H. Press, Astrophys. J.
{\bf 193}, 443 (1974); W. H. Press and S. A. Teukolsky, Astrophys.
J. {\bf 185}, 649 (1973).

\bibitem{StaChu} A. A. Starobinsky, Zh. Eksp. Teor. Fiz. {\bf 64},
48 (1973) [Sov. Phys. JETP {\bf 37}, 28 (1973)]; A. A. Starobinsky
and S. M. Churilov, Zh. Eksp. Teor. Fiz. {\bf 65}, 3 (1973) [Sov.
Phys. JETP {\bf 38}, 1 (1973)]

\bibitem{Det} S. Detweiler, Astrophys. J. {\bf 239}, 292 (1980).

\bibitem{Abram} M. Abramowitz and I. A. Stegun, {\it Handbook of
Mathematical Functions} (Dover Publications, New York, 1970).

\bibitem{Hodol} Note that ${\cal F}(\mu=1)=\sqrt{3}/2$, in which case Eq. (\ref{Eq19}) agrees
with the results obtained in \cite{Hod2}.

\bibitem{Hart} T. Hartman, W. Song, and A. Strominger, JHEP 1003:118
(2010); M. Cvetic and F. Larsen, JHEP 0909:088 (2009).

\bibitem{Noterp} Note that for KN black holes $r_{\pm}=M\pm(M^2-a^2-Q^2)^{1/2}$.

\end{thebibliography}
\end{document}